\documentclass[aps,twocolumn,showpacs,floatfix,byrevtex,titlepage,balancelastpage,raggedbottom,amsfonts,amssymb,amsmath]{revtex4}
\usepackage{epsfig}
\usepackage{bm}
\usepackage{times}
\usepackage{mathptmx}
\usepackage{amsmath}
\usepackage{graphicx}
\usepackage{graphics}
\usepackage{amssymb}
\usepackage[english]{babel}
\begin{document}
\title{Energy transport and fluctuations in small conductors}
\author{Danilo Sergi}
\affiliation{University of Applied Sciences (SUPSI),
The iCIMSI Research Institute, Galleria 2, CH-6928 Manno, Switzerland}
\date{\today}
\begin{abstract}
The Landauer-B\"uttiker formalism provides a simple and insightful way
for investigating many phenomena in mesoscopic physics. By this approach
we derive general formulas for the energy currents and apply them
to the basic setups. Of particular interest are the noise properties.
We show that energy current fluctuations can be induced by zero-point
fluctuations and we discuss the implications of this result.
\end{abstract}
\maketitle

\section{Introduction}
The study of quantum effects in small conductors is generally referred to
as mesoscopic physics. The wave nature of the electrons is relevant and
many counterintuitive results appear: the quantization of the conductance,
persistent currents in small loops, the quantum Hall effect
and the weak localization effect, to cite but a
few (for a review see Ref.~\cite{datta} and references therein).
In the Landauer-B\"uttiker formalism the motion of the electrons in
the conductors is described as a scattering process. This approach was
originally proposed to investigate the conductance of a single-channel
wire \cite{landauer1,landauer2}
and then extended to other structures \cite{ring,channels,terminals,hall}
and properties \cite{noise_1990,noise_1992,review,pretre}.
Instead, much less interest have received the thermoelectric properties.
The first
investigations of energy transport in mesoscopic conductors appeared in
Refs.~\cite{anderson,imry_thermo,streda}. The average properties were
studied also in Refs.~\cite{butcher,guttman}, while the noise properties in
Ref.~\cite{krive}. In this paper we extend the Landauer-B\"uttiker formalism to
account for energy transport and fluctuations. We derive the energy counterpart
of several results characterizing the electrical properties of mesoscopic 
conductors. The role of irreversible processes is at the center of our 
attention, especially in equilibrium at low temperatures.
In this regime we show that in a two-terminal
conductor energy exchange can happen, of course, under the constraint
of no net flow of energy.

\section{General results}
\textit{The model.} We consider a multi-terminal many channel coherent
conductor. This means that the energy carriers can enter or leave the sample
through $M$ leads with $N_{\alpha}$ transverse channels,
$\alpha=1,\ldots,M$, and their motion from one lead to another is
phase coherent. Each lead is connected to an electron reservoir
characterized by the temperature $T_{\alpha}$, the chemical
potential $\mu_{\alpha}$ and the Fermi-Dirac distribution function
$f_{\alpha}(E)=\{\exp((E-\mu_{\alpha})/k_{\mathrm{B}}T_{\alpha})+1 \}^{-1}$,
where $k_{\mathrm{B}}$ is the Boltzmann constant. The reservoirs absorb
all incident electrons irrespective of their phase and energy.
Furthermore, the reservoirs are incoherent, that is, the electrons
emerging from different reservoirs do not have any phase
relationship and their phase is also independent of that of
absorbed electrons. We neglect any interaction of the electrons
with other electrons or with phonons, magnetic impurities, et cetera.
At the conductor elastic scattering processes take place. The
elastic scattering properties of the conductor are described by
the scattering matrix $\bm{S}$. It relates the amplitude of the
outgoing states to the amplitude of the incoming states. Let
$\bm{S}_{\alpha\beta}(E)$ be the submatrix of dimension
$N_{\alpha}\times N_{\beta}$ defined as $\big(\bm{S}_{\alpha\beta}(E)\big)_{mn}
=S_{\alpha\beta,mn}(E)$, $m=1,\ldots,N_{\alpha}\ \text{and}\ n=1,
\ldots,N_{\beta}$. $\bm{S}_{\alpha\beta}(E)$ connects the incident
amplitudes in lead $\beta$ to the outgoing amplitudes in lead $\alpha$.
An energy carrier arriving at the conductor in contact $\beta$ in
channel $n$ has probability $R_{\beta\beta,mn}=\mid
S_{\beta\beta,mn}  \mid^{2}$ to be scattered back into contact
$\beta$ in channel $m$ and probability $T_{\alpha\beta,mn}=\mid
S_{\alpha\beta,mn}  \mid^{2}$ to be scattered into contact
$\alpha$ in channel $m$. Evidently, for a carrier in contact
$\beta$ the total probability of reflection and of transmission
into contact $\alpha$ are given by, respectively,
$R_{\beta\beta}=\sum_{mn}R_{\beta\beta,mn}=\sum_{mn}\mid
S_{\beta\beta,mn}
\mid^{2}=\mathrm{Tr}(\bm{S}^{\dag}_{\beta\beta}\bm{S}_{\beta\beta})$ and
$T_{\alpha\beta}=\sum_{mn}T_{\alpha\beta,mn}=\sum_{mn}\mid
S_{\alpha\beta,mn}
\mid^{2}=\mathrm{Tr}(\bm{S}^{\dag}_{\alpha\beta}\bm{S}_{\alpha\beta})$ ;
$\mathrm{Tr}$ stands for trace. The conservation of the energy carriers
imposes that $\bm{S}$ is unitary.

\textit{Average properties.} We assume that the energy  carriers are only
electrons and we do not take into account the spin degeneracy. The classical
expression of the energy current in lead $\alpha$ is given by
$W_{\alpha}(t)=(1/e)\int (E-\mu)\mathrm{d}I_{\alpha}(t,E)$.
At low temperatures the chemical potential $\mu$ can be
assumed to be approximately the energy value above which transport
occurs, i.e., the Fermi energy $E_{\mathrm{F}}$. We subtract it from the
total energy of the energy carriers since we are interested in the
net energy flowing through the leads, to which the Fermi sea does not
contribute. With calculations similar to those made in
Refs.~\cite{noise_1990,noise_1992,review,pretre},
we find that the average energy current in lead $\alpha$ is
\begin{equation}
\langle \hat W_{\alpha}(t)\rangle=\frac{1}{h}\sum_{\beta}\int \mathrm{d}E(E-\mu)
(N_{\alpha}\delta_{\alpha\beta}-T_{\alpha\beta}(E))f_{\beta}(E)\ ,
\label{w}
\end{equation}
where $\delta_{\alpha\beta}$ is the Kronecker delta.
We consider the linear response regime, i.e., for all $\beta$ we
write $\mu_{\beta}=E_{\mathrm{F}}+\Delta\mu_{\beta}$ and
$T_{\beta}=T+\Delta T_{\beta}$. $E_{\mathrm{F}}$ is the Fermi energy of
the electrons in the reservoirs and $T$ is approximately the
average temperature of the system.
When $\Delta\mu_{\alpha}$ and $\Delta T_{\alpha}$ are supposed to be small,
we find that we can write the above expression as
\begin{equation}
\langle \hat
W_{\alpha}(t)\rangle=\sum_{\beta}\Delta\mu_{\beta}K^{\Delta
\mu}_{\alpha\beta}+\sum_{\beta}\Delta T_{\beta}K^{\Delta
T}_{\alpha\beta}\ ,
\label{linear}
\end{equation}
with the thermal conductance matrices $K^{\Delta\mu}_{\alpha\beta}$ and
$K^{\Delta T}_{\alpha\beta}$ defined as
\begin{equation*}
K^{\Delta \mu}_{\alpha\beta}=\frac{1}{h}\int \mathrm{d}E(E-E_{\mathrm{F}})
\Big(-\frac{\partial f(E)}{\partial E}\Big)
\big(N_{\alpha}\delta_{\alpha\beta}-T_{\alpha\beta}(E) \big)
\end{equation*}
and
\begin{multline}
K^{\Delta T}_{\alpha\beta}=\frac{1}{hT}\int \mathrm{d}E (E-E_{\mathrm{F}})^{2}
\nonumber
\Big(-\frac{\partial f(E)}{\partial E}\Big)\times\\
\big( N_{\alpha}\delta_{\alpha\beta}-T_{\alpha\beta}(E) \big)\ ;
\nonumber
\end{multline}
$f(E)=\{\exp((E-E_{\mathrm{F}})/k_{\mathrm{B}}T)+1\}^{-1}$.
From Eq.~(\ref{linear}) we see that, in the linear response regime, there are
two, clearly independent, contributions to energy transport due to a temperature
or a chemical potential gradient. In the zero temperature limit, of course,
one has to use Eq.~(\ref{w}), as we shall see in the next section.

\textit{Fluctuations. }The spectral density of energy current fluctuations
$S_{\alpha\beta}^{W}$ is defined by
\begin{multline}
\nonumber
2\pi S_{\alpha\beta}^{W}(\omega)\delta(\omega+\omega')=\\
\langle \Delta\hat W_{\alpha}(\omega) \Delta\hat
W_{\beta}(\omega')+\Delta\hat W_{\beta}(\omega')\Delta\hat W_{\alpha}(\omega)
\rangle\ .
\nonumber
\end{multline}
We indicate by $\Delta\hat W_{\alpha}(\omega)=\hat
W_{\alpha}(\omega)-\langle\hat W_{\alpha}(\omega)\rangle$ the Fourier
transform of the fluctuating part of the energy current operator
in lead $\alpha$. We introduce the matrix \cite{noise_1992}
$\bm{A}_{\beta\gamma}(\alpha,E,E')=\mathbb{I}_{\alpha}\delta_{\alpha\beta}
\delta_{\alpha\gamma}-\bm{S}_{\alpha\beta}^{\dag}(E)\bm{S}_{\alpha\gamma}(E')$;
$\mathbb{I}_{\alpha}$ is the identity matrix $\alpha\times\alpha$.
By following closely the analysis proposed in
Refs.~\cite{noise_1990,noise_1992,review}, we find that
\begin{multline}
S_{\alpha\beta}^{W}(\omega)=\frac{1}{h}\int \mathrm{d}E \
(E+\frac{\hbar \omega}{2}-\mu)^{2}\times\\
\sum_{\delta\gamma}\mathrm{Tr}\big(\bm{A}_{\delta\gamma}(\alpha,E,E+\hbar
\omega)\bm{A}_{\gamma\delta}(\beta,E+\hbar \omega,E ) \big)\times\\
\big\{  f_{\delta}(E) [1-f_{\gamma}(E+\hbar \omega)]+f_{\gamma}(E+\hbar
\omega)[1-f_{\delta}(E)] \big \}\ .
\label{s}
\end{multline}
From the physical quantities entering this formula we see that
energy current noise is determined by the transmission properties
of the conductor and the statistics of the energy carriers. It is
straightforward to verify that our result satisfies
\begin{equation}
S_{\alpha\beta}^{W}(-\omega)=S_{\beta\alpha}^{W}(\omega)\ .
\label{u1}
\end{equation}
In the zero-frequency limit there is another useful identity. For the
unitarity of the scattering matrix we obtain
\begin{equation}
\sum_{\alpha}S_{\alpha\beta}^{W}(0)=\sum_{\beta}S_{\alpha\beta}^{W}(0)=0\ .
\label{u2}
\end{equation}

We conclude this section by making the remark that noise evokes the
idea of disorder, but $S_{\alpha\beta}^{W}$ is also a measure of the
correlation of the deviations away from the average value of the
energy current in the leads. Another important point is that noise
is determined by both the particle and the wave properties of the
energy carriers. Indeed, $S_{\alpha\beta}^{W}$ is derived starting from
operators in the second quantization formalism. The details can not be
given here exhaustively, but we address the reader to
Refs.~\cite{noise_1992,review} for analogous calculations.

\section{Applications}
\begin{figure}[b]
\begin{center}
\includegraphics[width=8.6cm]{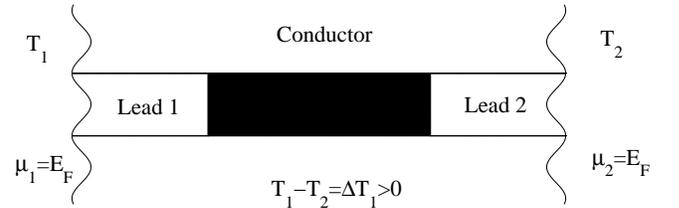}
\caption{Two-terminal conductor in the presence of a temperature gradient:
$T_{1}\neq T_{2}$.}
\label{fig1}
\end{center}
\end{figure}

\textit{The quantum of thermal conductance.} As a first application of
our general results, and notably of Eq.~(\ref{linear}), we consider a
two-terminal conductor.  The leads
have the same number of channels, $N$, and we suppose that energy
transport is due only to a temperature gradient, that is, $\Delta
\mu_{1}=\Delta \mu_{2}=0$, $\Delta T_{2}=0$ and $\Delta T_{1}\neq
0$, as shown in Figure \ref{fig1}. In the basis of
the eigen-channels and by assuming that the scattering matrix is
approximately constant over the energy range where transport
occurs, the energy current through the two leads is given by
\begin{multline}
W= \langle \hat W_{1}\rangle  =-\langle \hat W_{2}\rangle=
\Delta T_{1}\sum_{n}T_{n}(E_{\mathrm{F}})\times\\
\frac{1}{hT}\int \mathrm{d}E (E-E_{\mathrm{F}})^{2}\Big( -\frac{\partial
f(E)}{\partial E} \Big)\ .
\nonumber
\end{multline}
We denote $T_{n}(E_{\mathrm{F}})$ the eigenvalues of the matrix
$\textbf{S}^{\dag}_{21} \textbf{S}_{21}$ evaluated at the Fermi
energy. They should not be confused with the temperature $T$. At
low temperatures the integral in the above result can be
estimated. We obtain the quantum of thermal conductance
\cite{anderson,rego}:
\[
K_{\mathrm{o}}(T)=\frac{1}{hT}\int \mathrm{d}E(E-E_{\mathrm{F}})^{2}\Big( -\frac{\partial
f(E)}{\partial E} \Big)\cong \frac{\pi^{2}k_{\mathrm{B}}^{2}}{3h}T\ .
\]

If now we apply a small voltage across the conductor, we readily
obtain the Wiedemann-Franz law
$L=K_{\mathrm{o}}(T)/TG_{\mathrm{o}}=(\pi^{2}/3)(k_{\mathrm{B}}/e)^{2}$,
where $G_{\mathrm{o}}=e^{2}/h$ is the quantum of conductance. $L$ is
usually referred to as the Lorentz number.

\textit{Dissipation and non-equilibrium noise.} Let us consider a
two-terminal conductor
at zero temperature over which a small voltage $V$ is applied.
We choose $\Delta\mu_{1}=eV$ and $\Delta\mu_{2}=0$. The leads have
the same number of channels. Making use of the Landauer formula, which
yields the average current $I =(e^{2}/h)T_{12}V$,
and of the unitarity of the scattering matrix, from
Eq.~(\ref{w}) we readily obtain $\langle \hat
W_{1}\rangle=-\langle \hat W_{2}\rangle=(1/2)I V$.
We immediately see that $\langle \hat W_{1}\rangle+\langle \hat
W_{2}\rangle=0$, and thus the conductor does not absorb energy.
This result is usually interpreted as follows. We might write the
energy current flowing through the leads as
$(I/e)\times eV/2$. $I/e$ represents the flow
of particles through the conductor, and $eV/2$ is the average
excess energy of the electrons. When an electron enters the
sample, it leaves behind a hole with approximately the same
energy. In order to obtain the total energy dissipated in the
reservoirs we have also to take into account the energy released
by holes. This is done by multiplying the energy current in the
leads, to which contribute only the electrons, by a factor
of $2$. This yields the expected result $ I V$.
Nevertheless, this analogy with the ohmic behavior is only
formal. In mesoscopic conductors we have a spatial separation
between elastic and inelastic scattering. Our result of course
depends on the geometry of the conductor via its transmissive
behavior, but the energy is dissipated in the reservoirs.

Let us come to the energy current noise properties.
From Eq.~(\ref{s}), in the basis of eigen-channels, we obtain
\begin{equation}
S_{11}^{W}(0)=\frac{2}{3h}\sum_{n}T_{n}(1-T_{n})(eV)^{3}\ ,
\label{shot}
\end{equation}
and from Eqs.~(\ref{u1}) and (\ref{u2}) we see that
$S_{11}^{W}(0)=S_{22}^{W}(0)=-S_{12}^{W}(0)=-S_{21}^{W}(0)$.
For completeness, let us point out that in the low transparency
limit, i.e. $T_{n}\ll1$, corresponding for example to the case of
a tunnel barrier, we have
$S_{11}^{W}(0)=(2/3)(e^{3}/h)\sum_{n}T_{n}\
V^{3}=(2/3)e  I    V^{2}$,
where we have used Landauer formula $G=(e^{2}/h)\sum_{n}T_{n}$ for the
conductance, which yields the average current flowing through the conductor
$I =GV$. The above result is usually referred
to as the classical limit. It corresponds to the case where the
emission of electrons is uncorrelated and, as a result, the instants of
emission are random and governed by a distribution function of the
Poisson type \cite{review}.
\begin{figure}[t]
\begin{center}
\includegraphics[width=8.6cm]{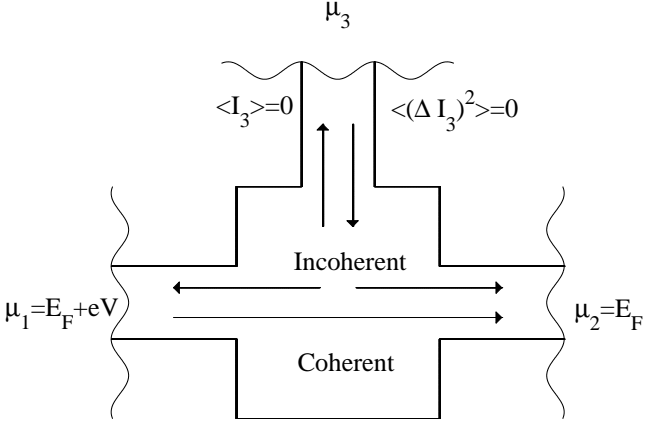}
\caption{Two-terminal conductor with a voltage probe. A fraction of the
 energy carriers are scattered coherently and the others incoherently in the
 forward and backward direction.}
\label{fig2}
\end{center}
\end{figure}

\textit{Inelastic scattering.}
We now study the effect of inelastic scattering on energy
transport. Within the scattering formalism, neglecting any kind of
interaction, it is possible to introduce inelastic scattering by
adding a fictitious voltage probe to the mesoscopic conductor
\cite{inelastic1}, as shown in Figure \ref{fig2}. This model for inelastic
scattering has the advantage of reducing the study of inelastic
scattering to an elastic scattering problem with the further
requirement of local current conservation at the voltage probe.
An ideal voltmeter has an infinite internal impedance and
therefore at the voltage probe the current vanishes at any moment
of time \cite{inelastic1,inelastic2}: $\langle I_{3} \rangle
=\langle (\Delta I_{3})^{2} \rangle=0$.
This means that when an electron is absorbed by the voltage probe
reservoir its phase and energy are randomized, and immediately
another electron is injected into the conductor with an energy
and a phase uncorrelated with those of the outgoing electron. The
energy current flowing through the conductor has both a coherent and an
incoherent component. A fraction of the electrons is scattered
coherently from contact $1$ to $2$ and the others are scattered
inelastically in the forward and in the backward direction.
We concentrate ourselves on the case of completely incoherent transmission,
i.e.~$T_{21}=T_{12}=0$, and thus $T_{3\alpha}=T_{\alpha 3}$, at zero temperature. 
By using Eq.~(\ref{w}) we find for the energy current in the three leads
\begin{eqnarray*}
\langle \hat{W}_{1} \rangle=\frac{T_{13}}{h}\frac{(eV)^{2}}{2}-\frac{1}{h}
\frac{T_{13}T_{31}^{2}}{(T_{31}+T_{32})^{2}}\frac{(eV)^{2}}{2}\ ,\\
\langle \hat{W}_{2} \rangle=-\frac{1}{h}\frac{T_{23}T_{31}^{2}}{(T_{31}+T_{32})^{2}}
\frac{(eV)^{2}}{2}\ ,\\
\langle \hat{W}_{3}\rangle=-\frac{1}{h}\frac{T_{31}T_{32}}{T_{31}+T_{32}}
\frac{(eV)^{2}}{2}\ .
\end{eqnarray*}
The unitary of the scattering matrix guarantees that $\langle \hat{W}_{1}\rangle
+\langle \hat{W}_{2}\rangle+\langle \hat{W}_{3}\rangle=0$, and so all dissipation
processes occur in the reservoirs. Then, the voltage probe
reservoir absorbs energy: the electrons entering
the voltage probe are thermalized through inelastic scattering
and release a fraction of their excess energy. $\langle \hat{W}_{3}\rangle$
is thus nothing but the Joule heat dissipated in the voltage probe 
(cf.~Refs.~\cite{inelastic1,ibm}).

We study instead  energy current fluctuations
in the quasi-elastic regime. This means that the electron
entering the voltage probe is replaced by an electron with the
same energy, but an uncorrelated phase \cite{jong}. This is the
reason why this model is generally employed to simulate
phase-breaking processes. Energy conservation is achieved by
demanding that at the voltage probe current is conserved in each
energy interval \cite{jong}.
It is worth noting that phase-breaking processes do not affect the average
energy current flowing through the conductor.
In fact, we find that $\langle \hat{W}_{1}\rangle=-\langle
\hat{W}_{2}\rangle= I V/2$, as obtained for the two-terminal conductor.
For the noise properties, from Eq.~(\ref{s}), in the zero-frequency limit,
we find that
\begin{multline}
S_{11}^{W}(0)=\frac{2}{3}e I
V^{2}\big[\frac{e^{2}}{h}\sum_{n}T^{(1)}_{n}(1-T^{(1)}_{n})R_{1}^{4}+\\
\frac{e^{2}}{h}\sum_{n}T^{(2)}_{n}(1-T^{(2)}_{n})R_{2}^{4} +
R_{1}^{2}R_{2}+R_{1}R_{2}^{2}\big]/R^{3}\ ,
\nonumber
\end{multline}
where $T_{n}^{(1)}$ and $T_{n}^{(2)}$ designate the transmission
probabilities from contact 1 to 3 and from contact 3 to 2,
respectively (see Fig.~\ref{fig2}); then,
$R=G^{-1}=R_{1}+R_{2}$ is the total resistance of the conductor, with
$R_{1}=(h/e^{2})/T_{31}$ and $R_{2}=(h/e^{2})/T_{32}$.
As before, $S_{11}^{W}(0)=S_{22}^{W}(0)=-S_{12}^{W}(0)=-S_{21}^{W}(0)$.
Interestingly, for a ballistic conductor the above result
does not vanish, in contrast to Eq.~(\ref{shot}), but reduces to
$S_{11}^{W}(0)=(2/3)e I
V^{2}R_{1}R_{2}(R_{1}+R_{2})^{-2}$.
This indicates that the presence of phase-breaking
processes are associated with energy current fluctuations.

\textit{Equilibrium noise.}
We recall that in equilibrium the power spectrum of current
fluctuations is given by
\begin{equation}
S^{I}(\omega)=4 G E(\omega,T)\ ,\quad\text{where}
\label{fdt}
\end{equation}
\begin{equation}
E(\omega,T)=\frac{\hbar \omega}{2}+\frac{\hbar \omega}{
\exp(\hbar \omega/k_{\mathrm{B}}T)-1}\ .
\label{ho}
\end{equation}
$G$ is the conductance of a two-terminal conductor and $E(\omega,T)$
is the average energy at temperature $T$ of an oscillator of
frequency $\omega$, being the sum of the zero-point energy and the
Planck spectrum. Equation (\ref{fdt}) is known as the
fluctuation-dissipation theorem, stating that equilibrium is
governed by irreversible processes at the microscopic level causing
fluctuations because the system experiences a fluctuating
force arising from the interaction with its environment \cite{nyquist,callen}.
At high temperatures Eq.~(\ref{ho}) reduces to the classical equipartition
value, indicating that the
fluctuating force originates from thermal agitation, while at low
temperatures we are left with the quantum of zero-point energy. We want to
understand whether vacuum fluctuations are associated with energy exchange.
Let us first consider energy current noise at a
non-vanishing temperature $T$ in the zero-frequency limit. A
simple calculation shows that Eq.~(\ref{s}) yields
$S_{\alpha\beta}^{W}(0)=2k_{\mathrm{B}}T^{2}\big(  K_{\alpha\beta}^{\Delta T}
+K_{\beta\alpha}^{\Delta T} \big)$.
This is the Johnson-Nyquist formula for energy current noise.
Now, at zero temperature we find that
\begin{equation}
S_{\alpha\beta}^{W}(\omega)=\frac{2}{3}\frac{1}{e^{2}}(G_{\alpha\beta}+
G_{\beta\alpha})\Big(\frac{\hbar\mid \omega  \mid}{2} \Big)^{3}\ .
\label{st0}
\end{equation}
For clarity we have written the result in terms of the
conductance matrix
$G_{\alpha\beta}=(e^{2}/h)(N_{\alpha}\delta_{\alpha\beta}-
T_{\alpha\beta})$. The only fundamental constant that enters this
result is the Planck constant, and we see that energy current
noise is proportional to $\hbar^{2}$, in line with what
obtained in Ref.~\cite{nagaev}. Equation (\ref{st0}) is the
main result of our work.
It is interesting to consider the case of a ballistic, single-channel,
two-terminal conductor because this situation admits a simple interpretation.
We find that $S_{11}^{W}(\omega)=S_{22}^{W}(\omega)=-S_{12}^{W}(\omega)=-S_{21}^{W}(\omega)=(\hbar^{2}/12\pi)\mid\omega\mid^{3}> 0$. This means that the energy current
fluctuates and if a mode tends to enter the sample in a lead, the same mode 
tends to leave the sample from the other lead. It also follows that energy 
transport is forbidden only on the average.

\section{Conclusions}
Within a unified framework we have investigated energy transport and
fluctuations in mesoscopic conductors. Importantly, our results on noise 
can be of relevance for the debate on dephasing from vacuum fluctuations
\cite{nagaev,webb,zaikin,cedraschi,imry_dephasing}. In the Landauer-B\"uttiker
formalism there are no fluctuating forces appearing explicitly, but
we neglect any kind of interaction in the leads. For this reason,
Eq.~(\ref{st0})
allows us to conclude that energy exchange between the reservoirs is
forbidden only on the average. Finally, the conductor and the leads
form a conservative hamiltonian system and ultimately we have shown
with an example that the coherence of an open quantum system is not
always fully preserved also in equilibrium at very low temperatures.

\begin{acknowledgments}
The author thanks Prof.~Markus B\"uttiker for the supervision
of the diploma thesis from which this article is obtained.
\end{acknowledgments}

\end{document}